\documentclass{emulateapj}
\def\xte{{\it RXTE~}} 

\shortauthors{Lin et al.}

\begin{document}

\title{Type I X-ray Bursts from the Neutron-star Transient XTE J1701-462}

\author{Dacheng Lin\altaffilmark{1}, Diego Altamirano\altaffilmark{2}, Jeroen Homan\altaffilmark{1}, Ronald A. Remillard\altaffilmark{1}, Rudy Wijnands\altaffilmark{2}, and Tomaso Belloni\altaffilmark{3}}
\altaffiltext{1}{MIT Kavli Institute for Astrophysics and Space Research, MIT, 70 Vassar Street, Cambridge, MA 02139-4307; lindc@mit.edu}
\altaffiltext{2}{Astronomical Institute, ``Anton Pannekoek'', University of Amsterdam, and Center for High Energy Astrophysics, Kruislaan 403, 1098 SJ Amsterdam, The Netherlands.}
\altaffiltext{3}{INAF -- Osservatorio Astronomico di Brera, Via  E. Bianchi 46, I-23807, Merate (LC), Italy}
\begin{abstract}
 
The neutron-star X-ray transient \object{XTE J1701-462} was observed
for $\sim$3 Ms with \xte during its 2006-2007 outburst. Here we report
on the discovery of three type-I X-ray bursts from \object{XTE
J1701-462}. They occurred as the source was in transition from the
typical Z-source behavior to the typical atoll-source behavior, at
$\sim10\%$ of the Eddington luminosity. The first burst was detected
in the Z-source flaring branch; the second in the vertex between the
flaring and normal branches; and the third in the atoll-source soft
state. The detection of the burst in the flaring branch cast doubts on
earlier speculations that the flaring branch is due to unstable
nuclear burning of accreted matter. The last two of the three bursts
show photospheric radius expansion, from which we estimate the
distance to the source to be 8.8 kpc with a 15$\%$ uncertainty. No
significant burst oscillations in the range 30 to 4000 Hz were found
during these three bursts.

\end{abstract}


\clearpage

\section{INTRODUCTION}
\label{sec:intro}
As accreted H/He-rich matter accumulates on the surface of a neutron
star (NS), it is compressed and heated, occasionally leading to
violent thermonuclear burning, a phenomenon known as type I X-ray
bursts. Such bursts were discovered during the mid-1970s
\citep{grgusc1976,becoev1976}, and many theoretical studies have been
made to investigate their detailed properties
\citep[e.g.,][]{jo1977,lala1978,nahe2003,wohecu2004,cona2006}.  For
reviews, see \citet{levata1993}, \citet{cu2004}, and
\citet{stbi2006}. Recently, \citet{gamuha2006} presented a large
sample of bursts (1187 in total) observed by \xte over a time interval
of more than ten years.

The bursts typically rise very rapidly ($<$2 s), followed by a slower
exponential decay ($\sim$10 s to several minutes). The burst spectra
can be fit with a single blackbody (BB)
\citep{swbebo1977,holedo1977,gamuha2006}. Burst properties do depend
on the mass accretion rate ($\dot{m}$) and the composition of the
accreted matter (See Table~1 by \citet{gamuha2006} and references
therein and above). Roughly, for solar metallicities, the bursts
mostly burn helium and have a fast rise when $\dot{m}$ is about
1--10$\%$ of the Eddington limit ($\dot{m}_{\rm Edd}$). At higher or
lower $\dot{m}$, they burn mixed hydrogen and helium and have a slower
rise and decay. If the accreted matter is dominated by helium, the
bursts mostly burn helium, and their properties vary little with the
accretion rate. The accreted matter is expected to burn stably at
$\dot{m}$ close to or above $\dot{m}_{\rm Edd}$ (although bursts are
seen in some Z sources, see below). In bright bursts, the photospheric
layer can be lifted off the NS surface by radiation pressure when the
local X-ray luminosity reaches the Eddington limit. Under certain
assumptions, these bursts can then be used as distance estimators
\citep{balesz1984,kudein2003,gapsch2003}.

Bursts are only found in weakly magnetized NS X-ray binaries, which
mainly consist of two classes, i.e., atoll and Z sources
\citep{hava,va2006}. They have different luminosity, spectral and
timing properties, etc. Atoll sources have $\dot{m}$ typically
$<0.5\dot{m}_{\rm Edd}$ and can have hard and soft states in addition
to transition between them. In the hard state, a significant fraction
of the energy is emitted above 20 keV and is normally accompanied by
radio jets \citep[e.g.,][]{fe2006,mife2006}. The source in the soft
state has relatively much weaker emission above 20 keV and is believed
to be dominated by emission from the accretion disk and the boundary
layer where the accreted matter impacts the NS surface. Bursts are
very often seen in atoll sources and have been observed in any of the
above X-ray states \citep[e.g.,][]{gamuha2006}.

Z sources include six classical objects \object{Sco X-1}, \object{GX
17+2}, \object{GX 349+2}, \object{Cyg X-2}, \object{GX 340+0}, and
\object{GX 5-1} \citep{hava}, with $\dot{m}$ close to or above
$\dot{m}_{\rm Edd}$. In the color-color diagram, Z sources can trace
out a 'Z' pattern, whose branches are called horizontal, normal and
flaring branches (HB/NB/FB), respectively. The Z pattern can move in
the diagram, most obviously in the case of \object{Cyg X-2}. Although
the spectra in Z sources are all soft and dominated by the emission
from the accretion disk and boundary layer, these branches are
different from each other in spectral and timing properties and
evolution timescales. Only \object{GX 17+2} has exhibited type I X-ray
bursts, and no bursts have been reported to be in the FB
\citep{kagr1984,tahiki1984,szvale1986,kuhova2002}. There are burst
events in \object{Cyg X-2}, but their thermonuclear origin is still
inconclusive \citep{kuvava1995,wivaku1997,gamuha2006}.

In this letter, we report an analysis of X-ray bursts from the NS
X-ray binary \object{XTE J1701-462} during its 2006-2007
outburst. \object{XTE J1701-462} is a unique source in that it is the
first NS transient that shows Z-source characteristics
\citep{hovawi2007,lireho2009}.  During the decay of the outburst, the
source changed from the typical Z-source behavior to the typical
atoll-source behavior \citep{lireho2009, hoetal2009}. The analyses of
comprehensive pointed observations of this outburst by \xte have
improved our understanding of atoll and Z sources and the physical
processes associated with the three Z-source branches. For example,
the spectral analysis in \citet{lireho2009} showed that the evolution
from the Z to atoll sources and the movement of the 'Z' tracks for Z
sources in the color-color diagram result from the change in
$\dot{m}$. The three Z-source branches were also suggested to
represent three distinct physical mechanisms presumably operating at
constant $\dot{m}$. Our analysis of its bursts is another part of our
campaign to understand this source.

Preliminary results of the bursts from \object{XTE J1701-462} have
been given in \citet{hobewi2007,howial2007} and \citet{lihore2007}. In
\S\ref{sec:reduction}, we first carry out a systematic search of the
\xte archive for bursts from this source. We then present the spectral
fit results of the bursts and estimate the distance of this source in
\S\ref{sec:spectralfit}. We describe our search for burst oscillations
in \S\ref{sec:osc}. Finally we discuss our results and summarize our
conclusions in \S\ref{sec:con}.

\section{OBSERVATIONS AND BURST SEARCH}
\label{sec:reduction}

\begin{deluxetable*}{llccc}
\tabletypesize{\scriptsize}
\tablecaption{Three bursts from XTE~J1701-462 observed by \xte\label{tbl-1}}
\tablewidth{0pt}
\tablehead{
& Burst number &  I & II &  III\\
&Observation ID           & 93703--01--01--01    & 93703--01--02--00  &  93703--01--02--08 
}
\startdata

Persistent& Count rate (cts\,s$^{-1}$\,PCU$^{-1}$)    & 396  & 301 & 176 \\
emission & Soft color               & 1.35   & 1.28 & 1.22 \\
&Hard color               &  0.46  & 0.44 & 0.49 \\
&Source branch/state\tablenotemark{a}  & FB &NB/FB vertex &atoll SS\\
&flux (2.5--25 keV, $10^{-9}$ erg\,cm$^{-2}$\,s$^{-1}$) & 4.45$\pm$0.07 & 3.38$\pm$0.04 & 1.99$\pm$0.02 \\
&flux (Bolometric, $10^{-9}$ erg\,cm$^{-2}$\,s$^{-1}$) & 6.17$\pm$0.12 & 4.90$\pm$0.10 & 2.94$\pm$0.05 \\
&Dimensionless flux $\gamma$ & 0.150$\pm$0.004 &0.119$\pm$0.003 & 0.072$\pm$0.002\\
\\
Burst\tablenotemark{b} &Start time (UT)   & 2007/07/17 12:24:22 & 2007/07/20 14:14:04  & 2007/07/25 13:39:24\\
& Rise time (s)    & 0.5 & 1.5 & 1.5 \\
& Decay time scale $\tau_1$ (s)  & 4.1$\pm$0.3 & 2.6$\pm$0.2 & 3.4$\pm$0.3 \\
&Decay time scale $\tau_2$ (s) & 10.6$\pm$5.0 & 5.6$\pm$1.9 & 8.5$\pm$4.7 \\
&Characteristic time scale $\tau$ (s) & 5.6$\pm$0.2 & 5.5$\pm$0.2 & 6.4$\pm$0.2\\
& Peak count rate (cts\,s$^{-1}$\,PCU$^{-1}$) & 2248 & 3382 &3465 \\
& Peak flux ($10^{-9}$ erg\,cm$^{-2}$\,s$^{-1}$)  & 27.2$\pm$1.0   &  39.2$\pm$1.1  &  42.9$\pm$1.2       \\
& Fluence ($10^{-7}$ erg\,cm$^{-2}$) & 1.52$\pm$0.03 &2.16$\pm$0.03 & 2.76$\pm$0.04\\
& Asymptotic radius (km, at 8.8kpc) &7.9$\pm$1.5 & 8.0$\pm$1.6 & 7.9 $\pm$1.3 \\
& Radius expansion?  & N & Y & Y \\
& Pulsed fraction upper limit & 37.9\%  & 14.6\%  &51.4\%
\enddata
\tablenotetext{a}{The source branch/state classification is from \citet{lireho2009}. FB: flaring branch; NB/FB vertex:
 the transition between the normal and flaring branches; atoll SS: atoll-source soft state}
\tablenotetext{b}{Persistent emission subtracted}
\end{deluxetable*}

\begin{figure}
\plotone{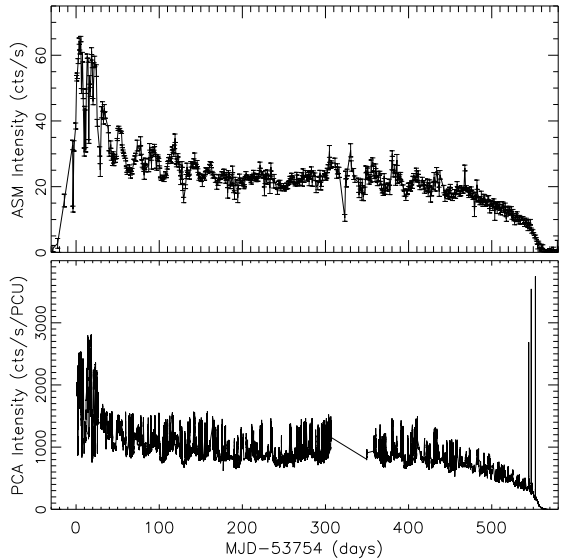} 
\caption{The one-day ASM light curve (upper panel) and 32-s PCA light
curve (lower panel). The three prominences in the decay of the PCA
light curve are type I X-ray bursts, where 1-s data are used.
\label{fig:lc}} 
\end{figure}

\begin{figure}
\plotone{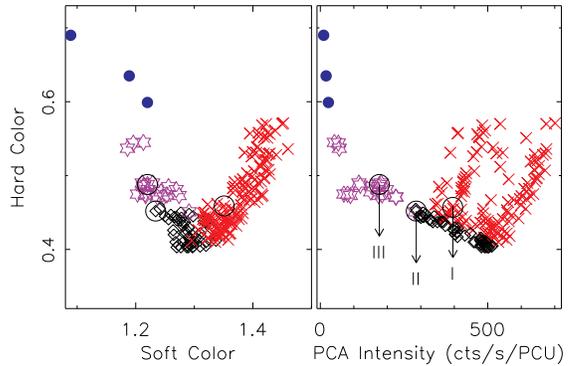} 
\caption{The color-color and hardness-intensity diagrams for
observations between MJD 54260 and 54315 in the decay of the 2006-2007
outburst of XTE~J1701-462. Blue filled circles for atoll-source hard
state, purple hexagrams for atoll-source soft state, red crosses for
Z-source flaring branch and black diamonds for the normal/flaring
branch vertex. The circled points mark the location of the bursts. The
data binsize is 960 s, but longer for observations in the atoll-source
soft and hard states (see text).
\label{fig:cd}} 
\end{figure}

\begin{figure}
\plotone{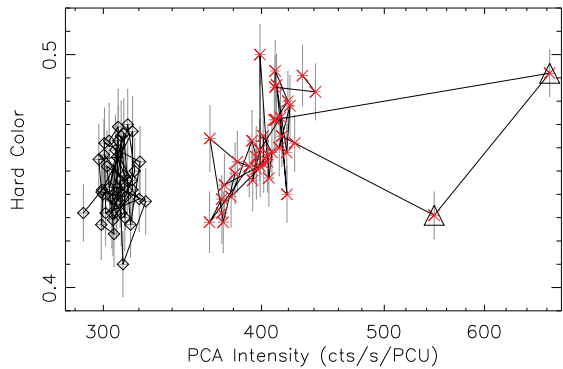} 
\caption{The hardness-intensity diagram for observations
93703--01--01--01 and 93703--01--01--08 using 32 s data. Black
diamonds are for the Z-source normal/flaring branch vertex and are all
from observation 93703--01--01--08, and red crosses for Z-source
flaring branch and all from observation 93703--01--01--01. The two
data points in the triangles are from the interval containing the
bursts. We connect the time-series data points with solid lines to
show the source evolution. The one--$\sigma$ error bars for the hard
color and intensity are also shown, with grey lines.
\label{fig:cdforfb}} 
\end{figure}

We carried out a systematic search for bursts from \object{XTE
J1701-462} during its 2006-2007 outburst. We used all 866 pointed
observations ($\sim$3 Ms) of \object{XTE J1701-462} made with the
Proportional Counter Array \citep[PCA;][]{jaswgi1996} on board {\it
RXTE}. We used the same scheme as that used by \citet{relico2006},
which is briefly described as follows. PCA 1-s light curves were
created for each observation from ``standard1'' data mode, which
integrates over the full energy range of the instrument (effectively
2--40 keV). For a time bin $t$, with intensity $C(t)$, we computed the
mean $b_1$ and the sample standard deviation $\sigma_1$ using 1-s data
in a ``background'' interval $t-280$ to $t-20$ and corresponding
values ($b_2$ and $\sigma_2$) in another ``background'' interval
$t+180$ to $t+280$ after it. Then we tested the joint condition
$[C(t)-b_1]>5\sigma_1$ and $[C(t)-b_2]>5\sigma_2$. If this condition
is satisfied for three sequential data points (at $t-1$, $t$ and
$t+1$), then we claim to find a burst candidate at $t$. Candidates are
rejected if they are due to detector breakdowns; see
http:$//$heasarc.gsfc.nasa.gov$/$docs$/$xte$/$recipes$/$pca$\_$breakdown.html
for more information.

Using the above scheme we find three burst candidates. They are on 2007
July 17, 20, and 25, during observations 93703--01--01--01,
93703--01--02--00, and 93703--01--02--08. We number them I, II, and
III, respectively (see Table~\ref{tbl-1} for details). All three
bursts have the above condition satisfied for $\sim$20 sequential data
points. The upper panel of Figure~\ref{fig:lc} shows the one-day light
curve from the All-Sky Monitor \citep[ASM;][]{lebrcu1996}, while the
lower panel shows the 32-s light curve from the Proportional Counter
Unit (PCU) 2 of the PCA, with 1-s bin size used for the three
observations with bursts detected to show these bursts more clearly.
The three bursts occur around the transition of \object{XTE J1701-462}
from the stage characteristic of a Z source to that of an atoll
source. 

Figure~\ref{fig:cd} shows the color-color diagram and
hardness-intensity diagram for observations from Modified Julian Day
(MJD) 54260 to 54315 in the decay of the outburst. The soft color and
hard color are the count rate ratios in the (3.6--5.0)/(2.2--3.6) keV
bands and the (8.6--18.0)/(5.0--8.6) keV bands, respectively. Each
data point has an integration time $\sim$960 s for observations before
MJD 54304, but corresponds to an entire observation for later
observations, which show little spectral variability. Several
observations are further combined when intensity is $<$30
counts\,s$^{-1}$\,PCU$^{-1}$. The figure is symbol-coded: blue filled
circles for atoll-source hard state, purple hexagrams for atoll-source
soft state, red crosses for Z-source FB and black diamonds for the
transition between the Z-source NB and the FB (NB/FB vertex). The
classification of the states/branches is from \citet{lireho2009}.

During the time interval for data shown in Figure~\ref{fig:cd}, the
source is not observed to enter the Z-source NB or HB. The FB in this
interval shows the same spectral evolution and timing properties as
the FB at higher luminosity when all Z-source branches observed
\citep{lireho2009,hoetal2009}. Given the shape of the color track
shown in Figure~\ref{fig:cd} and in view of the low persistent
luminosity ($\sim$10\% of $L_{\rm Edd}$, see below), one could also
classify the FB in Figure~\ref{fig:cd} as the atoll-source upper
banana branch. The evolution of the color tracks shown in
\citet{lireho2009} and \citet{hoetal2009} suggests that the
atoll-source upper banana branch and the Z-source FB as defined by
\citet{hava} describe a similar phenomenon, albeit at different
luminosities. In this paper, we will follow the state/branch
classification scheme by \citet{lireho2009}, in which the Z-source FB
phenomenon can be observed down to luminosities typically associated
with atoll sources.

In Figure~\ref{fig:cd}, we circle the data points where the bursts are
detected. Burst I is in the FB, burst II in the NB/FB vertex, and
burst III in the atoll-source soft state. The persistent emission
stays roughly constant for the observations when bursts II and III are
observed, and the occurences of the bursts seem not to modify the
persistent emission. We examine the same question in greater detail
for the observation containing burst I (Obs 93703--01--01--01), since
this is the first occurence of a burst from a Z source while it is in
the FB. In order to see whether the persistent emission is modified by
the burst, especially to see whether the source jumps to the NB/FB
vertex right after the burst, we show in Figure~\ref{fig:cdforfb} the
hardness-intensity diagram for this observation (red crosses). Results
for the next adjecent observation (Obs ID 93703--01--01--08; about one
day later), when the source is in the NB/FB vertex, are also shown
(black diamonds) for comparison. Data points with 32 s bin size are
used. The two data points in the triangles are from the interval
containing burst I. The data points are connected by black solid lines
in time to show the source evolution. We clearly see that the source
stays roughly at the same position in this diagram before and after
the burst. From this we conclude that this burst does not interfere
with the persistent emission in the FB either. Ramifications for this
result are considered in \S\ref{sec:con}.

The count rate, soft color, hard color, and flux of the persistent
emission are given in Table~\ref{tbl-1}. The flux is estimated using
a model of a multi-color disk plus a single-temperature blackbody (BB)
\citep{lireho2009}, with both 2.5--25 keV and bolometric values given
(unabsorbed). The bolometric flux is obtained analytically as our
model consists of two thermal components. The conversion of flux to
luminosity is complicated by uncertainties in several important
parameters, such as the source distance, the disk inclination and the
occultation factor of the BB, and is therefore not carried out
here. Instead, we give in Table~\ref{tbl-1} the dimensionless flux
\citep[$\gamma$;][]{vapele1988}, i.e., the ratio of the bolometric
persistent flux to the Eddington flux inferred from the radius
expansion bursts (see below). Based on this parameter, the $\dot{m}$
when the bursts are detected is roughly $10\%$ of $\dot{m}_{\rm Edd}$.

\section{BURST SPECTRAL FITS}
\label{sec:spectralfit}

\begin{figure*}\epsscale{1.0}
\plotone{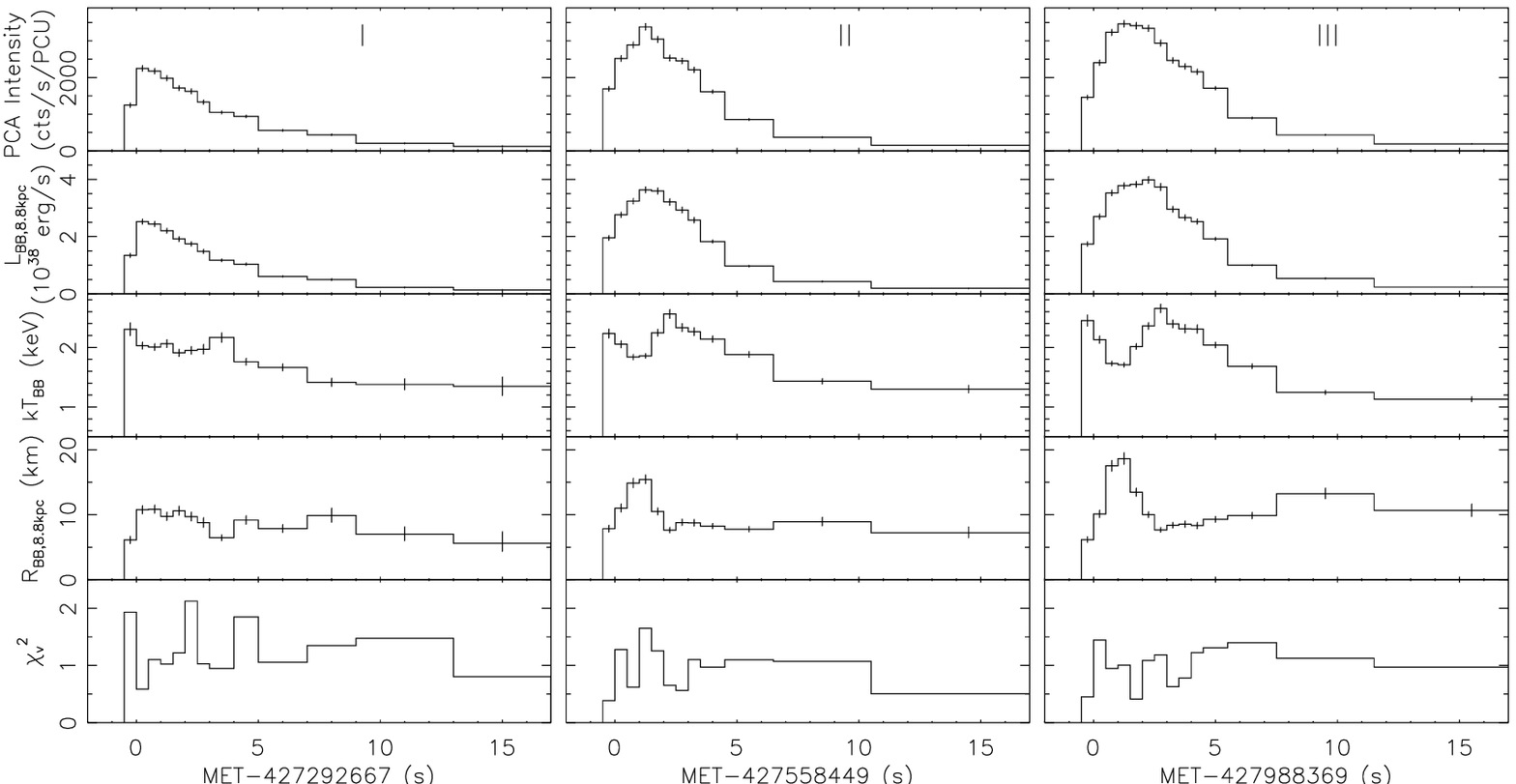}
\caption{The results of the spectral fits of time-resolved spectra of
the three bursts detected from XTE~J1701-462 during its 2006-2007
outburst. The latter two bursts show photospheric radius
expansion.
\label{fig:finalplot}} 
\end{figure*}

\begin{figure*} \epsscale{1.0}
\plotone{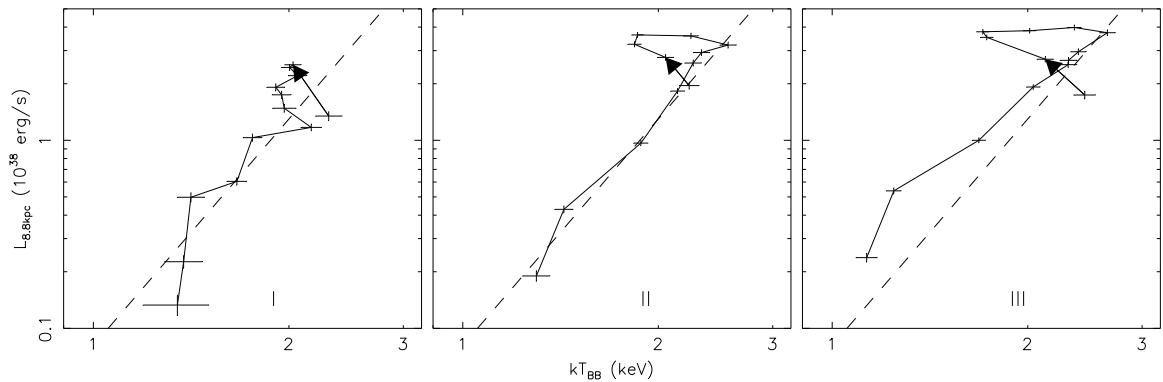}
\caption{The burst luminosity versus blackbody temperature. The arrows
show the direction of burst evolution. The dashed lines correspond to
$R_{\rm BB}=8$ km.
\label{fig:lumRT}} 
\end{figure*}

We used a standard procedure to do spectral fits of the three bursts:
create time-resolved spectra, use the persistent emission around the
burst as background, and fit the spectra with a BB model
\citep[e.g.,][]{gamuha2006}. We used a 15-s interval that ended 35 s
before the burst to define the persistent emission. Event mode data
with 64 channels and 125-$\mu s$ time resolution were used. Only PCUs
0 and 2 were on during the observations in which the above three
bursts were detected. Considering that the spectral calibration of PCU
0 has been bad at energies lower than 10 keV since the loss of the
propane layer in 2000, we only used data from PCU 2. We note that
\citet{lihore2007} used data from both PCUs. The peak count rates of
these bursts are quite high ($\sim$3500 counts/s/PCU), but by
comparing with the count rates from ``standard1'' mode data, we found
no telemetry saturation in the event mode data. The integration time
of the spectra was set to be 0.5 s around the peak and increased as
the count rate decreased to maintain sufficient statistics to
constrain the spectral parameters. An energy range of 3.5--20 keV was
used, and $0.8\%$ systematic errors were applied. Deadtime
corrections, which were as large as 5$\%$, were also made as suggested
by \xte team. The response files were created using HEAsoft version
6.4. A hydrogen column density $N_{\rm H}=2.0\times 10^{22}$ ${\rm
cm}^{-2}$ was assumed; it was inferred from spectral fitting of the
persistent emission using simultaneous observations of RXTE and
swift/XRT \citep{lireho2009}, and is consistent with the value
inferred from fitting of quiescence spectra \citep{fretal2009}.

The results of our spectral fits are shown in
Figure~\ref{fig:finalplot}.  There are five panels for each burst (all
with persistent emission background subtracted); PCA intensity,
bolometric luminosity, BB temperature $kT_{\rm BB}$, BB radius $R_{\rm
BB}$, and reduced $\chi^2$ are all shown versus time. The luminosity
and radius were calculated at a distance of 8.8 kpc (see below). The
low values of reduced $\chi^2$ suggest that the net burst spectra are
well fit by a BB. The fast rise and slow cooling decay confirm that
these bursts are bona fide type I X-ray bursts.

The most important feature that Figure~\ref{fig:finalplot} shows is
that the two brightest bursts (II and III) show photospheric radius
expansion, as indicated by the anti-correlation between the
temperature and radius when the luminosity reaches its peak: the
temperature decreases from $\sim$2.5 keV to $\sim$1.8 keV while the
radius increases from $\sim$10 km to $\sim$20 km. Radius expansion
bursts are also often studied in terms of the BB luminosity versus the
temperature, as shown in Figure~\ref{fig:lumRT}. In this figure, the
arrows indicate the direction of the temporal evolution. As shown in
the panels for bursts II and III, the source tends to evolve along a
horizontal line (constant luminosity) during the radius
expansion/contraction phase but along a (dashed) diagonal line
(constant radius) during the cooling decay phase.

We use these radius expansion bursts to estimate the distance to
\object{XTE J1701-462}. The peak flux in bursts II and III are given
in Table~\ref{tbl-1} with an average value of 41.1$\pm0.8 \times
10^{-9}$ erg\,cm$^{-2}$\,s$^{-1}$. Using the empirically determined
Eddington luminosity $3.79\pm0.15 \times 10^{38}$ erg\,s$^{-1}$ for
bursts showing photospheric radius expansion \citep[][with uncertainty
$15\%$]{kudein2003}, we derive a distance of 8.8 kpc. Using a
theoretical expression for the Eddington limit (see equation 8 in
\citet{gamuha2006}) and assuming a 1.4 solar-mass NS with a radius of
10 km, we obtain a distance of 7.3$\pm$0.1 kpc for H-poor case
(H-fraction $X=0$) and 5.6$\pm$0.1 kpc for H-rich case ($X=0.7$). We
assumed that the peak luminosity is reached when the photosphere has
settled back on the NS surface \citep{gamuha2006}, which is roughly
true in this case (Figure~\ref{fig:finalplot}). Hereafter, we use the
value derived by the empirical method, i.e., 8.8 kpc.

The other properties of the bursts are listed in Table~\ref{tbl-1}.
The start time was defined to be the time when the burst flux first
exceeded $25\%$ of the peak flux, and the rise time is the interval
that it takes for the burst flux to increase from $25\%$ to $90\%$ of
the peak value \citep{gamuha2006}. We see that all three bursts rise
very rapidly, i.e., within $\sim$1 s. Such bursts are probably due to
Helium burning, which is consistent with the $\dot{m}$ ($\sim$10$\%$
of $\dot{m}_{\rm Edd}$) during these bursts (\S\ref{sec:intro}). The
bursts show exponential decays but cannot be fit with a single
exponential curve. Thus, as in \citet{gamuha2006}, we divide the decay
(using the bolometric flux curve) into two parts and fit each with an
exponential curve with independent decay constants $\tau_1$ and
$\tau_2$. These bursts decay rapidly, on time scales of several
seconds. The fluence in Table~\ref{tbl-1} is estimated by summing the
fluxes over the burst and integrating the final exponential curve to
account for the additional flux beyond the data window. The
characteristic time scale $\tau$ is the fluence divided by the peak
flux. It is $\sim$6 s for all bursts.

As the bursts decay, the radius tends to remain at an asymptotic
value, as also shown in Figure~\ref{fig:lumRT}, and this is a signal
of unstable nuclear burning in the bursts spreading over the whole NS
surface \citep{levata1993}. We calculate the asymptotic radius in the
burst tails using data when the burst flux decays from $80\%$ to
$20\%$ of the peak value. The asymtotic radius for each burst is also
listed in Table~\ref{tbl-1} with an average of 8$\pm$1 km at a
distance of 8.8 kpc (corresponding to the dashed line in
Figure~\ref{fig:lumRT}). We note that there is a slight increase in
emission area in burst III as the decay progresses, as can be seen in
Figure~\ref{fig:lumRT}, the cause of which is unknown. The effects of
redshift and spectral hardening should be corrected for in order to
obtain the actual size of the NS \citep{levata1993}. If a 1.4
solar-mass NS and a hardening factor of 1.4 \citep[e.g.][]{majoro2004}
are assumed, then the actual size of the NS is $\sim13$ km.

\section{BURST OSCILLATION SEARCH}
\label{sec:osc}
We searched each burst for coherent pulsations in the frequency range
30 to 4000 Hz using Fourier techniques. We computed power spectra
throughout each burst using sliding 1, 2, 3 and 4-s windows with a
step of 0.125 s. The investigated energy bands were 2--60, 2--10, and
10--30 keV. We found no significant signal.  To estimate the upper
limits, we used the set of Leahy-normalized power spectra
\citep{ledael1983} with a 1-s window so as to take into account
possible frequency drifts.

The upper limit for each burst was then determined as follows
\citep[see also][]{vavawo1994}: (1) we searched for the largest
observed power $P_{max}$ in the 2--60 keV band; (2) we fit the noise
powers of the power spectrum in which we found $P_{max}$ with a
constant plus power law model; (3) we divided by the continuum model
and multiplied by 2 to re-normalize the power spectrum \citep[see,
e.g.,][]{isst1996,wast2006} and (4) we estimated the upper limits at a
99\% confidence level using the Groth distribution
\citep{gr1975,vavawo1994}. The pulsed fraction upper limits are listed
in Table~\ref{tbl-1}. They are relatively high and not very
constraining, compared with the typically observed amplitudes of
oscillations of a few ten percent \citep{stbi2006}.

\section{DISCUSSIONS AND CONCLUSIONS}
\label{sec:con}
We find three type I X-ray bursts from the 866 \xte pointed
observations ($\sim$3 Ms) of the NS transient \object{XTE J1701-462}
during its 2006-2007 outburst. These bursts are detected during the
decay of the outburst. The persistent emission of the observations
containing these three bursts is weak, with fluxes $\sim10\%$ of the
Eddington limit. Based on the source state/branch classifications by
\citet{lireho2009}, the first burst is detected in the Z-source
flaring branch; the second in the flaring/normal branch vertex; and
the third in the atoll-source soft state.

Detailed spectral fits confirm these bursts as type I X-ray
bursts. The last two of the three bursts show strong photospheric
radius expansion with a peak flux of 41.1$\pm0.8 \times 10^{-9}$
erg\,cm$^{-2}$\,s$^{-1}$. Assuming an Eddington luminosity of
$3.79\pm0.15 \times 10^{38}$ erg\,s$^{-1}$ \citep{kudein2003}, we
estimate a distance of 8.8 kpc for \object{XTE J1701-462} with $15\%$
systematic uncertainty. We find no significant burst oscillations
between 30--4000 Hz.

The distance to \object{XTE J1701-462} was initially estimated by
\citet{hovawi2007}. They gave a distance of $14.7\pm3$ kpc by
comparing the flux in the NB/FB vertex in a Sco-like Z-source interval
with that of \object{Sco X-1} (which has a distance of $2.8\pm0.3$
kpc, as determined by radio parallax measurements). This value is
higher than that derived here using radius expansion bursts. A
possible explanation for such a discrepancy is a difference in the
inclination of these two systems. The orbital inclination of
\object{Sco X-1} is $\sim$$40\degr$ \citep{fogebr2001,stca2002}. The
above discrepancy can be explained if \object{XTE J1701-462} has
orbital inclination of $70\degr$, as roughly estimated by
\citet{lireho2009} based on the weak iron emission lines detected,
assuming that the flux is dominated by the disk emission.

The occurrence of the burst near the end of the outburst is consistent
with the behavior of \object{XTE J1701-462} changing to that typical
of an atoll source. Atoll sources are known to be much more prolific
bursters than Z sources. Among the six classical Z sources, only
\object{GX 17+2} has exhibited type I X-ray bursts
(\S\ref{sec:intro}). These sources accrete at high $\dot{m}$, close to
or above $\dot{m}_{\rm Edd}$, as inferred from its radius expansion
bursts. \object{GX 17+2} has both short ($\tau\lesssim 10$ s) and long
($\tau>100$ s) bursts. Burst-like events from \object{Cyg X-2} are
very short ($\tau\sim3$ s), and many of them do not show cooling
during burst decay. Thus their thermonuclear origin is inconclusive
\citep{kuvava1995,wivaku1997,gamuha2006}. \object{XTE J1701-462} is
different in that its bursts are detected when it accretes at low
$\dot{m}$, $\sim$10$\%$ of $\dot{m}_{\rm Edd}$ and that all bursts are
short ($\tau\sim6$ s). From the $\dot{m}$ and the burst profiles (fast
rise and short duration), these bursts are consistent with helium
burning. Such bursts are often seen in the atoll sources in their
banana branch \citep{gamuha2006}. Thus the unstable nuclear burning in
these bursts should fall under the framework of the classical burst
theory, indicating that Z sources can have regular bursts. The results
that these bursts are consistent with helium burning can also be
explained if the NS is accreting H-poor material, as is often seen in
ultra-compact X-ray binaries, which have orbital periods of an hour or
less and have helium accreted from a degenerate donor. There is no
optical spectrum of the mass donor in \object{XTE J1701-462} to
determine whether it is H-poor. However, the 2006--2007 outburst is
very long ($\sim$20 months), while ultra-compact X-ray binaries have
never displayed such long outbursts when their light curves can be
associated with the class of soft X-ray transients. Thus the mass
donor in XTE J1701-462 is probably a normal star transferring H-rich
material.

So far no bursts have been detected in the FB (except in the NB/FB
vertex) from \object{GX 17+2} \citep{kuhova2002}, while the
thermonuclear origin of the Burst-like events from \object{Cyg X-2} is
inconclusive (see references above). While the detection of burst I
from \object{XTE J1701-462} in a low luminosity FB might not be
unique, given the abundance of burst in the atoll upper banana branch
\citep{gamuha2006}, it does cast doubt on the speculation that the
Z-source FB is caused by unstable nuclear burning
\citep{chhaba2006}. In \S\ref{sec:reduction}, we show that burst I
does not interfere with the FB behavior, in that no jump in the
hardness-intensity diagram is seen before and after the burst. In
\S\ref{sec:spectralfit}, we also show that burst I should have spread
over the whole NS surface, since there is constant emission area
during the decay. Thus, considering that the FB and bursts are
different (e.g., bursts rise in seconds, but the FB rises in minutes
or longer), if the FB is caused by unstable nuclear burning, it must
be a different type of nuclear burning from that for bursts, and it
can coexist when burst I burns nuclear material globally in
\object{XTE J1701-462}. Although we cannot simply exclude a complex
environment allowing the above scenario, current theoretic studies do
not favor coexistence of two types of unstable nuclear burning
\citep[e.g.,][]{hecuwo2007}. Thus the explanation of unstable nuclear
burning for the FB is quite uncertain. \citet{lireho2009} offered an
additional argument that the FB cannot be due to unstable nuclear
burning, as the energy released to produce the FB would require a much
higher $\dot{m}$ than is observed. \citet{lireho2009} propose an
alternative explanation for the FB, in which the inner disk radius
contracts temporarily from an Eddington-expanded value toward the
innermost stable circular orbit.

The burst rate depends on $\dot{m}$ and is expected to be zero at
$\dot{m}$ close to $\dot{m}_{\rm Edd}$ \citep{relico2006,gamuha2006}.
While this is consistent with the fact that we do not observe bursts
during the bright phase of the outburst for \object{XTE J1701-462} and
in most of Z sources, it does raise the question of why the Z source
GX 17+2 shows type-I X-ray bursts. One possible explanation could be
that bursts are possible at high $\dot{m}$, but only in the narrow
$\dot{m}$-range occupied by GX 17+2. We test this hypothesis by
estimating the expected number of bursts from \object{XTE J1701-462},
for the duration when it was similar to \object{GX 17+2}. We estimate
that \object{XTE J1701-462} has $\sim$0.5-Ms of exposure when it
resembles \object{GX 17+2}, i.e., with all three Sco-like Z branches
traced out, and thus has a similar $\dot{m}$ as well. However,
\object{XTE J1701-462} has no burst detected in this $\sim$0.5-Ms
interval, while we would have expected to see $\sim$6 bursts, based on
the burst rate of \object{GX 17+2} (12 bursts observed in 1 Ms with
\xte \citep{gamuha2006}). Thus the burst rate in \object{GX 17+2} is
much higher than \object{XTE J1701-462} at similar $\dot{m}$. It
suggests that bursts in \object{GX 17+2} are not only due to its
specific $\dot{m}$, but might also depend on other factors such as the
NS parameters or the chemical composition of the accreted matter.

The three bursts are detected at $\dot{m}$ $\sim10\%$ $\dot{m}_{\rm
Edd}$, with the burst rate about one per ten hours, considering that
the source is observed for several tens of hours around such
$\dot{m}$. This is consistent with the average value, $\sim$12 hours
from \citet{gamuha2006} or $\sim$6 hours from \citet{relico2006},
corresponding to the above $\dot{m}$.
 
This research has made use of data obtained from the High Energy
Astrophysics Science Archive Research Center (HEASARC), provided by
NASA's Goddard Space Flight Center. TB acknowledges support from ASI
through grant I/088/06/0.

\end{document}